\documentclass[12pt,preprint]{aastex}

\usepackage{graphicx}
\usepackage{epsfig,amsmath}

\newcommand{\be}{\begin{eqnarray}}
\newcommand{\ee}{\end{eqnarray}}

\begin{document}

\title{Long Term Cooling of Magnetar Crusts}

\author{David Eichler\altaffilmark{1}, Yuri Lyubarsky\altaffilmark{1},
Chryssa Kouveliotou\altaffilmark{2}, Colleen A.
Wilson\altaffilmark{2}} \altaffiltext{1}{Physics Department,
Ben-Gurion University, Be'er-Sheva 84105, Israel;
eichler@bgu.ac.il}
 \altaffiltext{2}{Marshall Space Flight Center,
Huntsville, Alabama}

\begin{abstract}
X-ray emission following giant flares of magnetars can be
categorized into three categories of time scales (a) short term
afterglow (b) medium term afterglow and (c) long term afterglow.
Short term afterglow, which declines over several hours,  seems to
correspond to gravitational resettling of uplifted material.
Medium term afterglow, which declines over several weeks or
months, appears to be the cooling of the heated outer crust, and
long term afterglow, which declines over a period of many years,
can be understood to be the cooling of the inner crust. The long
term afterglow profile may be a very sensitive indicator of
neutron star mass.
\end{abstract}

\maketitle

\section{Introduction}
The cooling of neutron stars is a very old subject; it dates back
almost to the discovery of  neutron stars.  Most of the cooling
theory has been directed at long term cooling following the
supernova that produces the neutron star. Recent reviews on the
subject include  Yakovlev  and Pethick (2004).  Eichler and Cheng
(1989) proposed that sufficiently energetic episodic energy
releases ($\ge 10^{43}$ ergs) in the crust might be able to heat
it enough to generate detectable {\it transient} X-ray emission
from the crust after such events as the crust cools. For many
years it appeared unlikely that episodic releases of $10^{43}$
ergs  or more could be arranged in the crusts of neutron stars.
The March 5, 1979 event was an outstanding exception, but at the
time it was one of a kind and mysterious.
Now, however, following the spin down measurements of Kouveliotou
et al. (1998, 1999) and the August 27, 1998 giant flare from SGR
1900+14, it is generally accepted that magnetars   (Duncan and
Thompson 1992, Thompson and Duncan, 1995, 1996), whose magnetic
fields contains $10^{47}$ to $10^{48}$ ergs, are able to release
sufficiently large amounts from their reserves of magnetic energy
to account for the giant flares. Indeed the three recorded such
events - the giant flares of  March 5, 1979, August 27, 1998, and
December 27, 2004 -  all seem to be similar and to be associated
with large magnetic reconnection powerful enough to move and heat
the crust of the magnetar.

The details of the giant flares are not the subject of this paper.
Rather, the intent is to sketch how  the X-ray afterglow seen from
the magnetars in the wake of these giant flares may teach us
something about the makeup of the magnetar crust. In particular,
it is significant that we can explain these x-ray light curves in
terms of normal neutron star material, where "normal" takes into
account the fact that the magnetic field is exceedingly large.
Checking the theory of normal matter in superstrong magnetic
fields  (e.g. by  Yakovlev and coworkers, see Potekhin 1999;
Yakovlev, Levenfish \& Shibanov 1999, Yakovlev et al 2001) against
observations sets a background against which  {\it abnormal}
neutron stars, such as those containing strange matter, could be
so identified.

Observations of outer crust cooling following intense heating may
also provide a rare opportunity to observe the effects, via the
data, of nuclear processes associated with strongly heated heavy
nuclei (SHHN) that are not observable in other situations, e.g.
reversible nuclear dissociation of, radioactivity of, pycnonuclear
activity in, and enhancement of neutrino cooling processes by
SHHN.

Several cautionary remarks should be issued from the start: First,
there is little doubt  that magnetospheric activity can cause
short term enhancement of the X-ray emission. The April 2001 flare
in SGR 1900+14 was a clear example of this. It produced a large,
short spike in the X-ray emission that stands out clearly in the
light curve (see Figure 3). Thus, we do not attempt to fit the
irregular bumps that appear on top of the overall long term
declining X-ray light curve.

Similarly, it is extremely likely that magnetospheric activity can
cause long term X-ray emission. The high persistent luminosity
from SGR 0526-66, (Rothschild, Kulkarni and Lingenfelter, 1994,
Kulkarni et al. 2003), about $1 \times 10^{36}$ erg/s, is too
large to be thermal emission from the surface 20 years after a
heating event, unless there is a huge amplification in the
magnetosphere. Some sort of energy release must be continuous, and
dissipation of magnetospheric currents seems the most likely. It
could, in fact, be argued that magnetospheric emission is
responsible for {\it all long term} X-ray emission from magnetars,
including the component that declines over several years. The high
pulsed fraction observed in the X-ray emission of many AXP's
certainly suggests that the pulsed emission comes from a heated
polar cap, and the amount is certainly comparable to what is
theoretically expected from current dissipation in the twisted
magenetic loops of the magnetospheres. We have no way of
disproving this alternative hypothesis at present. However, we now
have two cases - SGR 1627-41 after a period of high activity in
1998 and SGR 1900+14 after the giant flare of Aug 27, 1998 - for
which a curious pattern in the X-ray light curve was observed over
several years: a decline, followed by a plateau, followed, with
relative suddenness, by a steeper decline, and finally a leveling
out well below the level of the plateau. It is not clear why
$d^2L/dt^2$, the sudden change in the rate of decline of
luminosity L,  would be so high in a magnetospheric current
dissipation model for the X-ray decline. In the crustal cooling
model we have proposed (Kouveliotou et al 2003), there is a
natural explanation for the plateaus in terms of the heat storing
capacity of the inner crust.

Because magnetars are observed only occasionally, we do not have
detailed histories of the long term decline in the X-ray emission
following intense activity. The inferred monotonic decrease in
this emission over the several years following an active period is
generally based on only five or six points so far, and there is
always a small probability that the observed "pattern" is a matter
of chance.  This could be remedied with more frequent
observations.

Finally, the non-thermal spectra typical of the persistent X-ray
emission from SGR's suggest that the magnetosphere at least plays
some role, possibly cyclotron  upscattering of thermal seed
photons (Thompson, Lyutikov and Kulkarni 2002). The decline would
then be attributed to that of the seed photons in the crustal
cooling model, rather than to a decline in the extent of cyclotron
upscattering. The latter would predict softening of the spectrum
as the luminosity declined and this can be tested. There is an
alternative possibility that the non-thermal emission is from a
tenuous atmosphere heated by coronal currents (Thompson and
Beloborodov 2005; Beloborodov and Thompson 2006). This emission is
totally decoupled from the crustal cooling because, near the
surface of the neutron star, the cyclotron energy is too high.

{\it Short term afterglow}  arises when energy released in the
outer $10^{10}$ g/cm$^{2}$ of the crust lifts these layers above
the surface (Eichler and Cheng 1989, Eichler et al 2003). The
uplifted material settles in the strong gravitational field of the
magnetar, as the pairs whose pressure supports it against gravity
annihilate and release the heat, over the course of several hours,
through the top of the raised material layer. It is not hard to
show that in this case the luminosity decreases approximately as
$t^{-0.5}$ with a slight steepening near the end of the burst
(Eichler  et al 2003). This is in extremely good agreement ({\it
ibid}) with observation of short term afterglow from SGR 1900+14 (
Ibrahim et al 2001) after an SGR event on August 29, 1998 (shortly
after the giant flare). Because the fit is so sensitive to opacity
(and hence to the surface field) the data provide a rather
sensitive if indirect measurement of the surface magnetic field.
The best fit to the August 29, 1998 event (Eichler et al 2003) is
with a surface field of $6 \times 10^{14}$G, in excellent
agreement with the spindown estimate (Kouveliotou  et al. 1998).

In the {\it medium term afterglow} that was observed following the
August 27, 1998 flare in SGR 1900+14, the X-ray emission declined
as a power law over 40 days following the giant flare. Beyond that
point, the decaying X-ray tail merged with  (what was then
considered) the persistent X-ray emission of that magnetar (Woods
et al 2001). This 40 day afterglow was explained as the cooling of
the outer crust under the assumption that the energy released per
unit volume below the surface was comparable to that above the
surface (Lyubarsky, Eichler \& Thompson 2002). A fit is shown in
Figure 1 for a neutron star mass of 1.5 solar masses.

{\it Long term afterglow} was observed following a very active
period of bursting activity from the magnetar SGR 1627-41. Here
the X-ray luminosity declined gradually for about one year; it
then \textit{plateaued}  for the second year and dropped off very
sharply during the third year. It has since leveled out into what
we believe is probably pulsed emission from a polar cap that was
heated by the persistent current.  Kouveliotou et al. (2003)
explained this curious time profile in the cooling curve as being
due to the cooling of the inner crust of the magnetar. The plateau
phase during days 400-800 of this light curve can be explained as
being due to the huge heat storing capacity of the inner crust
which keeps the outer crust at a more or less constant temperature
profile until this reservoir of heat is used up. Once the heat is
used up, the surface cools very rapidly as observed. A good fit to
the data was obtained under the assumption that the inner crust
was heated to hundreds of millions of degrees but that the core
remained cool.  In order for the inner crust to cool within a year
or so, the surface gravity of the neutron star must be quite high,
at least 1.5  solar masses, and this implies that the core is
cooled by the direct URCA process (Gnedin et al. 2001). Thus the
assumption of a cool core is justified once the assumption of a
massive neutron star is made. Most of the heat in the inner crust
is conducted down  into the core and the duration of the plateau
in the surface emission lasts  only as long as the heat in the
inner crust remains. The fit from Kouveliotou et al (2003) is
reproduced and the data updated in Figure 2. The final leveling
out is attributed to magnetospheric current dissipation taking
over the main role of surface heating after the crust below has
cooled sufficiently. It is conjectured that during the decline of
the X-ray emission after a major heating event in 1998, SGR
1627-41 has become more like an AXP.

 The   X-ray emission from SGR 1900+14 beyond the first
 40 days after the August 1998 flare at first appeared to be the
persistent x-ray emission generated by heat stored in the core of
a young neutron star. However, eight years after this event it now
appears that this emission is actually in slow decline. It has
declined by a factor of two in the past eight years.  Nevertheless
we have been able to fit this decline with cooling of the inner
crust of SGR 1900+14 if we assume a high mass, $M=1.5 M_{\odot}$.
The medium term afterglow must now be fit with the same large
mass, which is more than originally assumed in the original paper
(Lyubarsky, Eichler and Thompson (2002), and in Figure 1 we have
replotted this fit for the higher mass.

Note that the cooling time profile is extremely sensitive not only
to neutron star mass via surface gravity and crust thickness, but
also sensitive to the assumed inner core temperature. This is due
in part to the non-linearities in the heat conduction.  Cool
material at these densities conducts heat much more effectively.
An abrupt end to a plateau in the cooling curve can happen only if
the core is cool, and this probably requires core cooling by the
direct URCA process. Data points  beyond those shown here, to
appear soon (Wilson et al, in preparation), should be able to
discriminate between the various theoretical curves plotted in
Figure 3.


In conclusion, the long term X-ray emission of the soft gamma
repeaters SGR 1627-41 and SGR 1900+14 following intense heating
episodes,  which declines  over a timescale of several years,  can
be modeled as cooling of their inner crusts under the assumption
that they are  made of otherwise normal material and that their
masses are at least $1.5 M_{\odot}$. It might be supposed that any
observed light curve could be fit by picking the initial
temperature distribution (ITD) that reproduces it.  However, it
turns out that there does not seem to be  {\it any}
ITD in the
crust that explains these light curves unless the mass is chosen
to be large and the core to be cool. If the core were hot - as
would be expected if the direct URCA process were suppressed - we
find no ITD that can fit the data without implausibly fine tuning.
It thus seems that observations of these long term cooling curves
may provide a good way to narrowly constrain the masses of neutron
stars that undergo large transient heating events.

Because the cooling time of the inner crust is such a
sensitive
function of the "margin of safety" - i.e. the difference
between the magnetar mass and the maximum mass -  it is in
principle possible to obtain some information about the clustering
of neutron stars near their maximum masses and  perhaps even the
fraction of them that become black holes. It would be interesting
to compare the distribution obtained in this way to that of binary
neutron star masses.

The authors gratefully acknowledge support from the Israel-US
Binational Science Foundation, the Israel Science Foundation, the
German-Israel Foundation, the Arnow Chair of Theoretical
Astrophysics, and the Israeli Ministry of Absorption.

\begin{figure*}
\includegraphics[width=8 cm,scale=0.6]{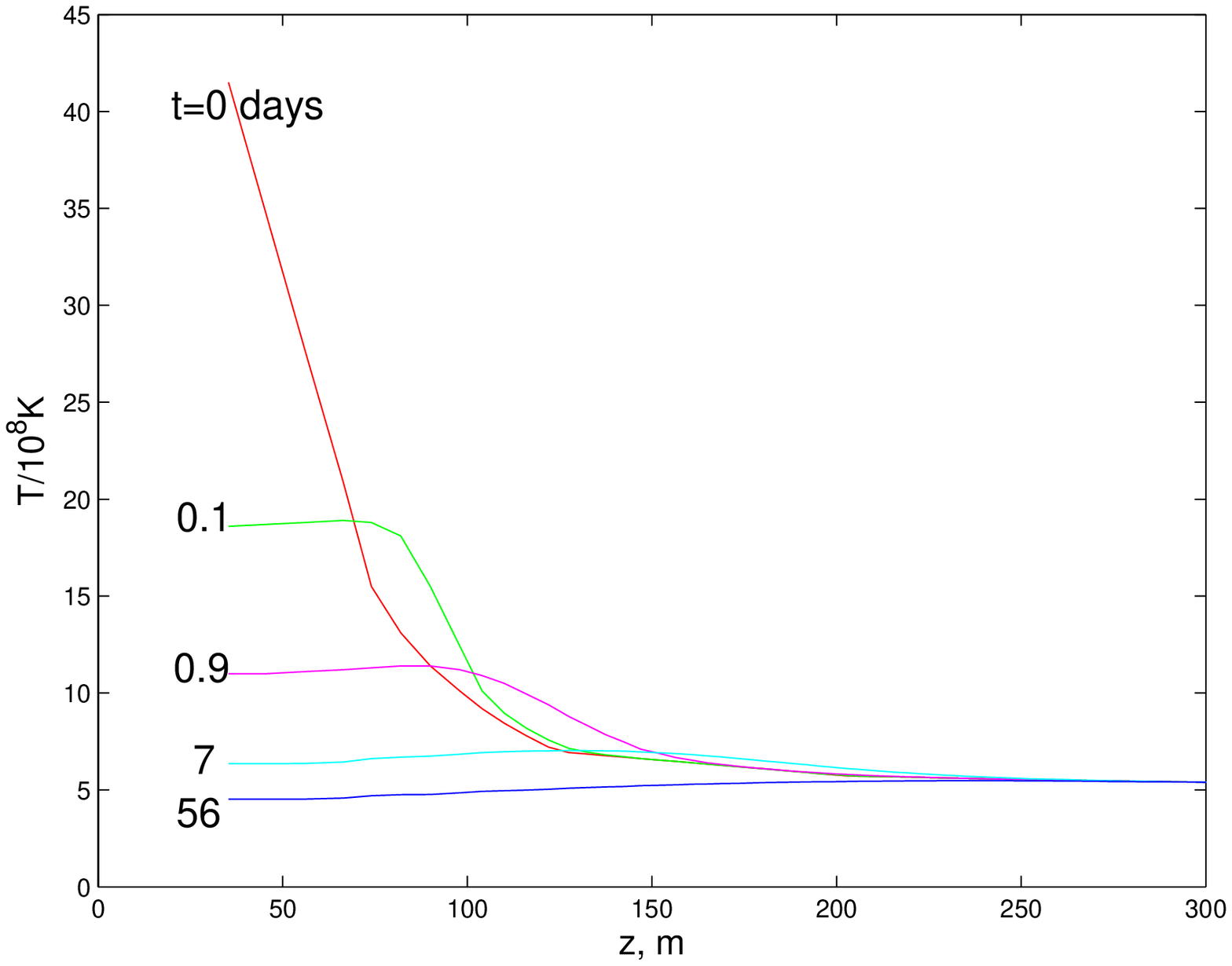}
\includegraphics[width=8 cm,scale=0.6]{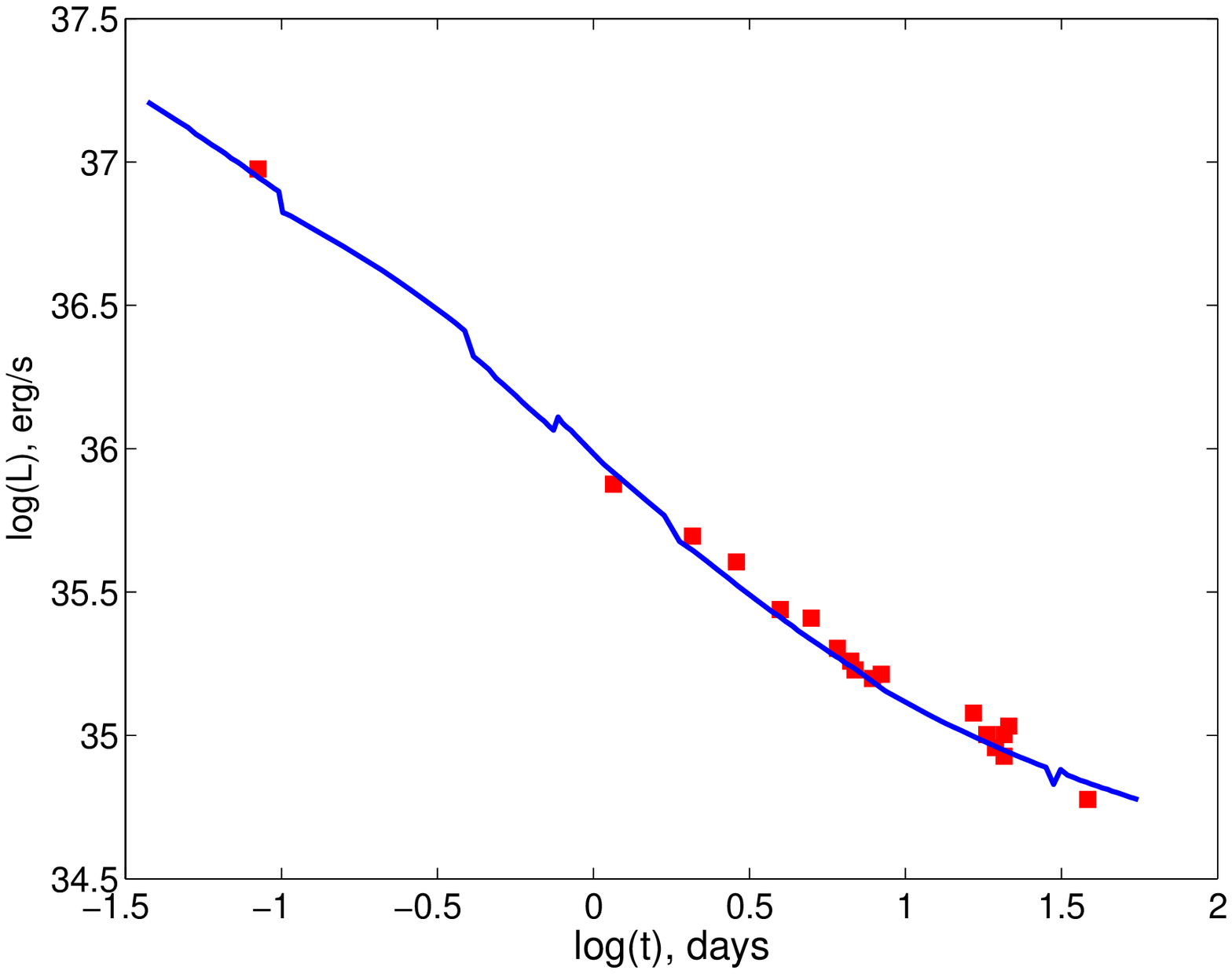}
\caption{Simulations of the medium-term afterglow from SGR
1900+14; a) evolution of the temperature distribution within the
upper crust with time $t$, days after the burst; b) luminosity,
squares show observational data }
\end{figure*}

\begin{figure*}
\includegraphics[scale=0.4]{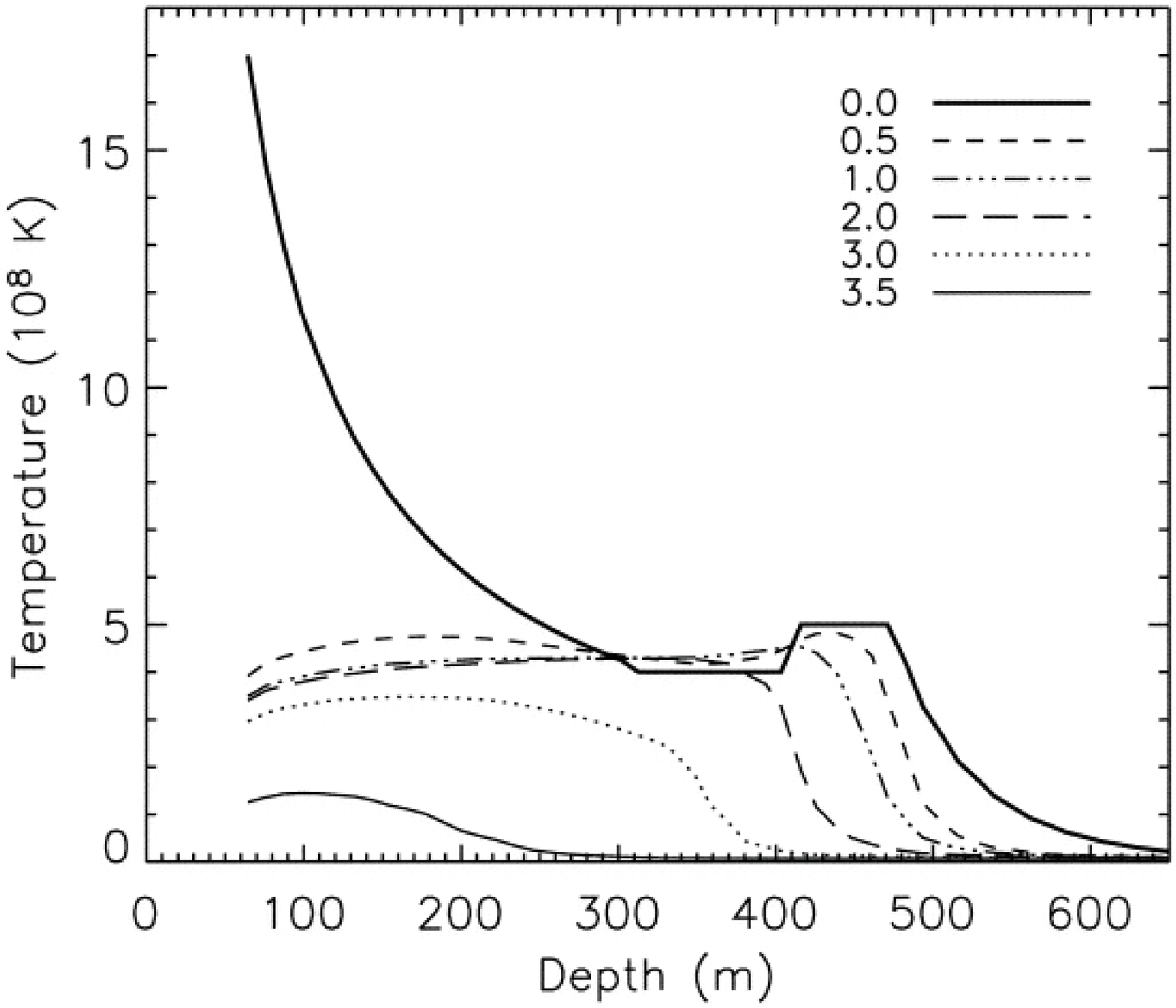}
\includegraphics[scale=0.4]{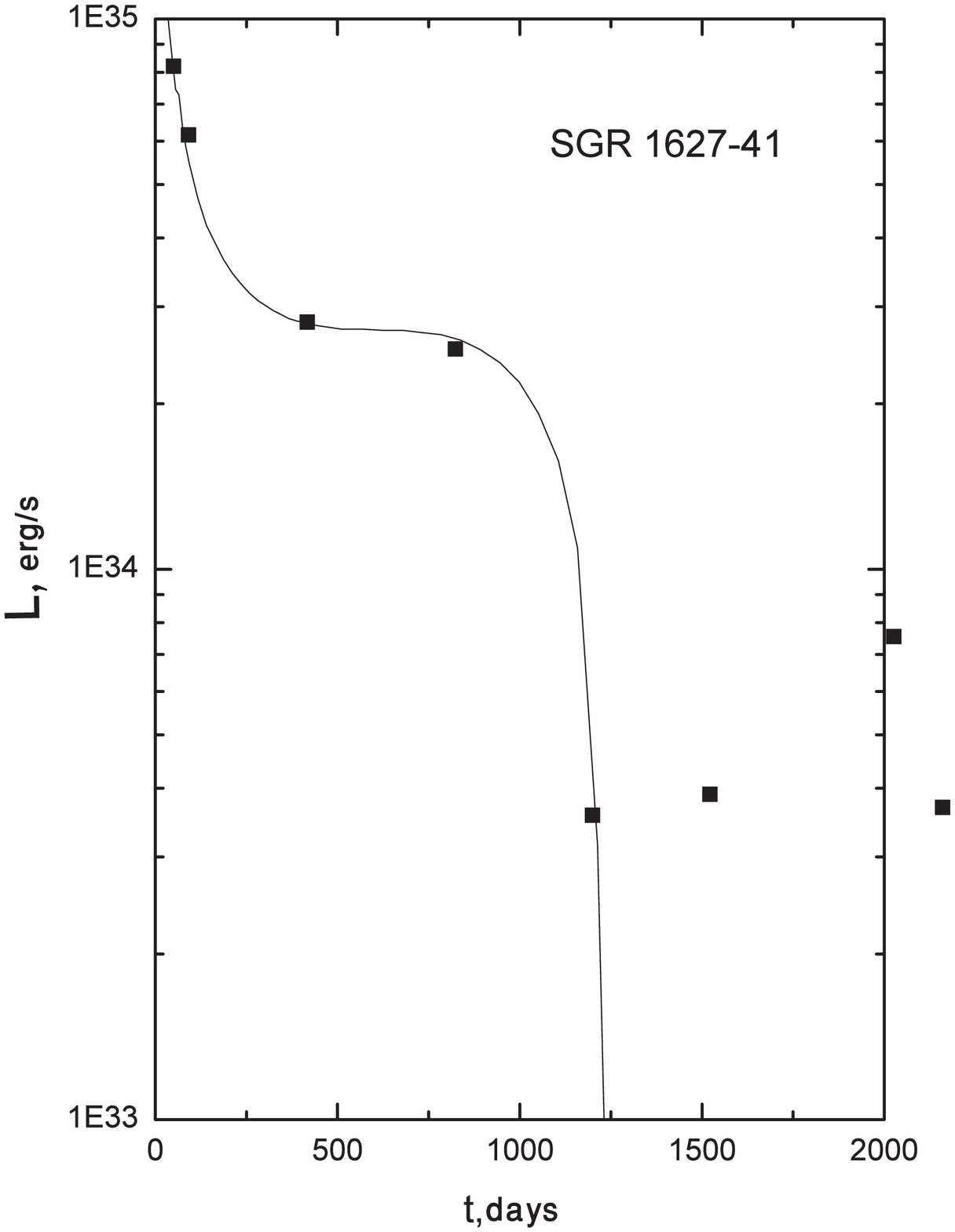}
\caption{Simulations of the long-term afterglow from SGR 1627-41
(Kouveliotou et al. 2003); a) evolution of the temperature
distribution within the crust; the different temperature profiles
correspond to years after the burst, as indicated in the plot
insert; b) luminosity, squares show observational data }
\end{figure*}

\begin{figure*}
\includegraphics[width=14 cm,scale=1.0]{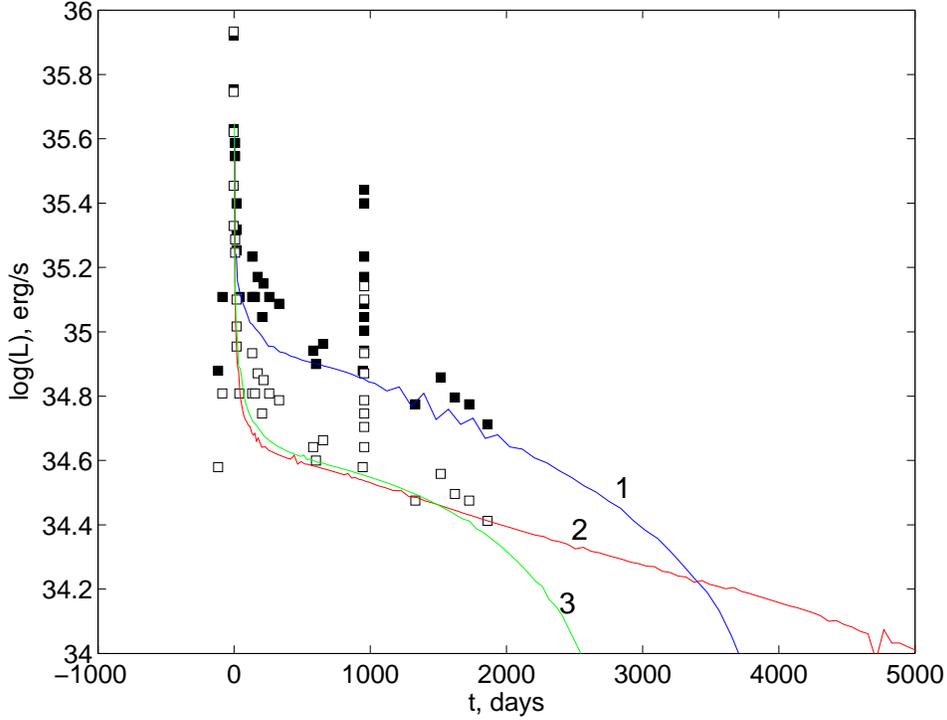}
\caption{Simulations of the long-term afterglow from SGR 1900+14.
Initial temperature distribution is qualitatively similar to that
in Fig. 2a: the temperature decreases with depth in the upper
crust, remains constant in the upper layers of the inner crust and
then sharply decreases until reaching the core temperature. The
most important parameters are the temperature of the inner crust,
$T_{inner}$, and that of the core, $T_{core}$. $T_{inner}=7\times
10^8$ K in curve 1 and $T_{inner}=5\times 10^8$ in curves 2 and 3.
$T_{core}=10^7$ K in curves 1 and 3 and $T_{core}=10^8$ K in curve
2. Empty squares show observational data for the distance 6 kpc,
filled squares for 9 kpc. }
\end{figure*}

\end{document}